\RequirePackage{fix-cm}
\documentclass{svjour3}                     
\smartqed  
\usepackage{graphicx}
\usepackage{amsmath}
\begin{document}

\title{Optimal joint remote state preparation of equatorial states
\thanks{X.L. is supported by the National Natural Science Foundation
of China under Grant No. 11004258 and the Fundamental
Research Funds for the Central Universities under Grant No.
CQDXWL-2012-014. S.G. acknowledges support from the
Ontario Ministry of Research and Innovation and the Natural
Sciences and Engineering Research Council of Canada.}}


\author{  Xihan Li        \and
        Shohini Ghose \and
}


\institute{
              Xihan Li \at Department of Physics, Chongqing University, Chongqing, China \\
              Department of Physics and Computer Science, Wilfrid Laurier University, Waterloo, Canada\\
              \email{xihanlicqu@gmail.com
              \and
              Shohini Ghose \at Department of Physics and Computer Science, Wilfrid Laurier University, Waterloo, Canada\\
              Institute for Quantum Computing, University of Waterloo, Canada, N2L 3G1}           
}

\date{Received: date / Accepted: date}

\maketitle

\begin{abstract}
We present a scheme for optimal joint remote state  preparation of two-qubit equatorial states. Our protocol improves on a previous scheme  (B. S. Choudhury and A. Dhara 2015 \emph{Quantum Inf. Process.} \textbf{14} 373) that had a success probability of 25\%, which increased to 50\% when extra classical information is sent to the receiver. We show that using our modified scheme, the desired state can be prepared deterministically with the same quantum channel. Moreover, we generalize the scheme to prepare $N$-qubit equatorial states in which the receiver can reconstruct the original state with 100\% success probability.

\end{abstract}

\section{Introduction}
Remote state preparation (RSP) \cite{RSP_lo}, like quantum teleportation \cite{tele},  is a novel way to transmit a quantum state between distant parties without physically sending the state itself.  Although it is only applicable to known states, RSP requires less classical communication than quantum teleportation \cite{RSP_Bennet}. Moreover, the two resources for quantum communication -  quantum entanglement and classical communication - can be traded off against each other in RSP schemes. Due to its interesting properties, RSP has been widely studied theoretically \cite{RSP_pati,low_entanglement_RSP,optimal_RSP,generalized_RSP,oblivious_RSP,faithful_RSP,cv_RSP} and experimentally \cite{RSPE1,RSPE2,RSPE3,RSPE4,RSPE5,RSPE6} in recent years.

To satisfy the requirements of different communication scenarios, RSP has several variants, one of which is called joint remote state preparation (JRSP) \cite{JRSP_I_GC1,JRSP_I_GC2}. In JRSP, the knowledge of the state to be prepared is shared by several senders, each of them having partial information. The receiver has no information about the state. Only when all the senders collaborate can the receiver reconstruct the desired state via some operations on his/her own particles. Many novel JRSP schemes have been designed for different types of quantum states using a variety of quantum channels \cite{JRSP_I_GC1,JRSP_I_GC2,AN0,JRSP1,JRSP2,JRSP3,AN1,AN2,AN3,Xiao,Peng}. Most recently, a JRSP scheme for preparing two-qubit equatorial states was proposed by Choudhury and Dhara \cite{JRSP2E}. In this scheme, two senders each have partial information about the state to be prepared and the quantum channel is composed of two maximally entangled three-qubit Greenberger-Horne-Zeilinger (GHZ) states. After the two senders apply projective measurements on their qubits and transmit their measurement outcomes, the receiver can reconstruct the original state with a success probability of 25\%. The authors also showed that the success probability can be increased to 50\% if one sender transmits extra classical information to the receiver. We henceforth refer to this scheme as the CD protocol.

In this letter, we revisit the scenario for JRSP of 2-qubit equatorial states explored in Ref. \cite{JRSP2E} and show that the CD protocol is not optimal. Our analysis demonstrates that by modifying the measurement basis of the two senders, the receiver can deterministically obtain the desired state with proper unitary operations. Moreover, we extend the scheme to JRSP of $N$-qubit equatorial states and explicitly describe the general form of the senders' measurement bases. In our scheme, the receiver can always perform a unitary operation corresponding to every possible measurement outcome of the senders and reconstruct the state with 100\% success probability.

\section{JRSP of an arbitrary equatorial two-qubit state}
There are three spatially separated parties in this JRSP scheme, the two senders Alice and Bob and the receiver Charlie. Alice and Bob wish to help Charlie prepare an arbitrary equatorial two-qubit state written as
\begin{eqnarray}
  \vert \Phi\rangle=\frac{1}{2}(\vert 00\rangle+e^{i \delta_1}\vert 01\rangle+e^{i \delta_2}\vert 10\rangle+e^{i \delta_3}\vert 11\rangle).
\end{eqnarray}   %
Here $\delta_j(j=1,2,3)$ is a real phase parameter shared by the two senders. Alice and Bob have partial information about this state. They each know the parameters $a_j$ and $b_j$, respectively, such that
\begin{eqnarray}
  \delta_j=a_j+b_j, (j=1,2,3).
\end{eqnarray}
The receiver has no knowledge about the desired state at all. Only when the two senders collaborate will the receiver be able to reconstruct  the two-qubit equatorial state in his location.

The quantum channel is composed of two maximally entangled three-qubit GHZ states.
\begin{eqnarray}
  &\vert& \Psi\rangle=\vert GHZ_3\rangle_{A_1B_1C_1}\otimes \vert GHZ_3\rangle_{A_2B_2C_2},\\
  &\vert& GHZ_3\rangle=\frac{1}{\sqrt{2}}(\vert 000\rangle+\vert 111\rangle).
\end{eqnarray}
Particles $A_j$ belong to Alice, and $B_j$, $C_j$ belong to Bob and Charlie, respectively. To help Charlie prepare the desired state, Alice and Bob perform projective measurements on their own qubits based on the partial information they have. The measurement basis for Alice and Bob is selected to be
\begin{eqnarray}
  \vert \varphi_0\rangle&=& \frac{1}{2}(\vert 00\rangle+e^{-ix_1}\vert 01\rangle+e^{-ix_2}\vert 10\rangle+e^{-ix_3}\vert 11\rangle),\\
   \vert \varphi_1\rangle&=& \frac{1}{2}(\vert 00\rangle+ie^{-ix_1}\vert 01\rangle-e^{-ix_2}\vert 10\rangle-ie^{-ix_3}\vert 11\rangle),\\
    \vert \varphi_2\rangle&=& \frac{1}{2}(\vert 00\rangle-e^{-ix_1}\vert 01\rangle+e^{-ix_2}\vert 10\rangle-e^{-ix_3}\vert 11\rangle),\\
     \vert \varphi_3\rangle&=& \frac{1}{2}(\vert 00\rangle-ie^{-ix_1}\vert 01\rangle-e^{-ix_2}\vert 10\rangle+ie^{-ix_3}\vert 11\rangle).
\end{eqnarray}
Here $x=a(b)$ for Alice (Bob).  The four states are mutually orthogonal and Alice (Bob) can obtain each one of them with equal probability. The quantum channel can be rewritten in terms of Alice's and Bob's measurement basis as
\begin{eqnarray}
  \vert \Psi\rangle= \frac{1}{8}&&[\vert\varphi_0\rangle_{A_1A_2}\vert \varphi_0\rangle_{B_1B_2}(\vert 00\rangle+e^{i \delta_1}\vert 01\rangle+e^{i \delta_2}\vert 10\rangle+e^{i \delta_3}\vert 11\rangle)_{C_1C_2}\nonumber\\
  &&+\vert\varphi_0\rangle_{A_1A_2}\vert \varphi_1\rangle_{B_1B_2}(\vert 00\rangle-ie^{i \delta_1}\vert 01\rangle-e^{i \delta_2}\vert 10\rangle+ie^{i \delta_3}\vert 11\rangle)_{C_1C_2}\nonumber\\
  &&+\vert\varphi_0\rangle_{A_1A_2}\vert \varphi_2\rangle_{B_1B_2}(\vert 00\rangle-e^{i \delta_1}\vert 01\rangle+e^{i \delta_2}\vert 10\rangle-e^{i \delta_3}\vert 11\rangle)_{C_1C_2}\nonumber\\
  &&+\vert\varphi_0\rangle_{A_1A_2}\vert \varphi_3\rangle_{B_1B_2}(\vert 00\rangle+ie^{i \delta_1}\vert 01\rangle-e^{i \delta_2}\vert 10\rangle-ie^{i \delta_3}\vert 11\rangle)_{C_1C_2}\nonumber\\
  &&+\vert\varphi_1\rangle_{A_1A_2}\vert \varphi_0\rangle_{B_1B_2}(\vert 00\rangle-ie^{i \delta_1}\vert 01\rangle-e^{i \delta_2}\vert 10\rangle+ie^{i \delta_3}\vert 11\rangle)_{C_1C_2}\nonumber\\
  &&+\vert\varphi_1\rangle_{A_1A_2}\vert \varphi_1\rangle_{B_1B_2}(\vert 00\rangle-e^{i \delta_1}\vert 01\rangle+e^{i \delta_2}\vert 10\rangle-e^{i \delta_3}\vert 11\rangle)_{C_1C_2}\nonumber\\
  &&+\vert\varphi_1\rangle_{A_1A_2}\vert \varphi_2\rangle_{B_1B_2}(\vert 00\rangle+ie^{i \delta_1}\vert 01\rangle-e^{i \delta_2}\vert 10\rangle-ie^{i \delta_3}\vert 11\rangle)_{C_1C_2}\nonumber\\
  &&+\vert\varphi_1\rangle_{A_1A_2}\vert \varphi_3\rangle_{B_1B_2}(\vert 00\rangle+e^{i \delta_1}\vert 01\rangle+e^{i \delta_2}\vert 10\rangle+e^{i \delta_3}\vert 11\rangle)_{C_1C_2}\nonumber\\
  &&+\vert\varphi_2\rangle_{A_1A_2}\vert \varphi_0\rangle_{B_1B_2}(\vert 00\rangle-e^{i \delta_1}\vert 01\rangle+e^{i \delta_2}\vert 10\rangle-e^{i \delta_3}\vert 11\rangle)_{C_1C_2}\nonumber\\
  &&+\vert\varphi_2\rangle_{A_1A_2}\vert \varphi_1\rangle_{B_1B_2}(\vert 00\rangle+ie^{i \delta_1}\vert 01\rangle-e^{i \delta_2}\vert 10\rangle-ie^{i \delta_3}\vert 11\rangle)_{C_1C_2}\nonumber\\
  &&+\vert\varphi_2\rangle_{A_1A_2}\vert \varphi_2\rangle_{B_1B_2}(\vert 00\rangle+e^{i \delta_1}\vert 01\rangle+e^{i \delta_2}\vert 10\rangle+e^{i \delta_3}\vert 11\rangle)_{C_1C_2}\nonumber\\
  &&+\vert\varphi_2\rangle_{A_1A_2}\vert \varphi_3\rangle_{B_1B_2}(\vert 00\rangle-ie^{i \delta_1}\vert 01\rangle-e^{i \delta_2}\vert 10\rangle+ie^{i \delta_3}\vert 11\rangle)_{C_1C_2}\nonumber\\
  &&+\vert\varphi_3\rangle_{A_1A_2}\vert \varphi_0\rangle_{B_1B_2}(\vert 00\rangle+ie^{i \delta_1}\vert 01\rangle-e^{i \delta_2}\vert 10\rangle-ie^{i \delta_3}\vert 11\rangle)_{C_1C_2}\nonumber\\
  &&+\vert\varphi_3\rangle_{A_1A_2}\vert \varphi_1\rangle_{B_1B_2}(\vert 00\rangle+e^{i \delta_1}\vert 01\rangle+e^{i \delta_2}\vert 10\rangle+e^{i \delta_3}\vert 11\rangle)_{C_1C_2}\nonumber\\
  &&+\vert\varphi_3\rangle_{A_1A_2}\vert \varphi_2\rangle_{B_1B_2}(\vert 00\rangle-ie^{i \delta_1}\vert 01\rangle-e^{i \delta_2}\vert 10\rangle+ie^{i \delta_3}\vert 11\rangle)_{C_1C_2}\nonumber\\
  &&+\vert\varphi_3\rangle_{A_1A_2}\vert \varphi_3\rangle_{B_1B_2}(\vert 00\rangle-e^{i \delta_1}\vert 01\rangle+e^{i \delta_2}\vert 10\rangle-e^{i \delta_3}\vert 11\rangle)_{C_1C_2}.
\end{eqnarray}
From this expression it is clear that no matter what measurement results Alice and Bob get, the state of $C_1C_2$ can always be transformed to Eq.(1) via a unitary operation. For example, if Alice and Bob's measurement results are $\vert \varphi_3\rangle_{A_1A_2}$ and $\vert \varphi_3\rangle_{B_1B_2}$, the unitary operation for $C_1C_2$ is $U=\vert 00\rangle\langle 00\vert-\vert 01\rangle\langle 01\vert+\vert 10\rangle\langle 10\vert-\vert 11\rangle\langle 11\vert=(I)_{C_1}\otimes (\sigma_z)_{C_2}$. Only when the measurement outcomes of both Alice and Bob are sent to Charlie can he prepare the desired state. The success probability of this JRSP scheme is 100\% in principle.

\section{JRSP of an arbitrary equatorial $N$-qubit state}
We now generalize our scheme to describe the deterministic JRSP of an arbitrary equatorial $N$-qubit state, which can be written as
\begin{eqnarray}
  \vert \Phi\rangle=\frac{1}{(\sqrt{2})^N}\sum^1_{l_1,...,l_N=0}\exp(i\delta_{j})\vert l_N,l_{N-1},...,l_1\rangle.
\end{eqnarray}
Here $j=0,1,2,...,2^N-1$, which is the decimal form of the binary string $(l_N,l_{N-1},...,l_1)$.
\begin{eqnarray}
  j=\sum^N_{n=1} 2^{n-1}l_n.
\end{eqnarray}
The two senders share partial information about the desired state with $a_j+b_j=\delta_j$ $(j=0,1,2,...2^N-1)$. The three parties share $N$ three-qubit GHZ states in advance.
\begin{eqnarray}
  \vert GHZ_3\rangle_{A_nB_nC_n}=\frac{1}{\sqrt{2}}(\vert 000\rangle+\vert 111\rangle)_{A_nB_nC_n}, (n=1,2,...,N).
\end{eqnarray}
Alice and Bob perform projective measurements on her/his particles $A_1,A_2,...,A_N$ and $B_1,B_2,...,B_N$, respectively. Their measurement basis can be written as
\begin{eqnarray}
  \vert \varphi_k\rangle=\frac{1}{(\sqrt{2})^N}\sum^1_{l_1,...,l_N=0}\exp(\frac{2\pi ijk-ix_j}{4})\vert l_N,l_{N-1},...,l_1\rangle,
\end{eqnarray}
where $k=0,1,2,...,2^N-1$ and $x=a(b)$ for Alice (Bob). Then the original $N$ GHZ states can be rewritten in terms of Alice's and Bob's measurement bases as
\begin{eqnarray}
  \vert \Psi\rangle&=&\vert GHZ_3\rangle_{A_1B_1C_1}\otimes\vert GHZ_3\rangle_{A_2B_2C_2}\otimes...\otimes\vert GHZ_3\rangle_{A_NB_NC_N}\nonumber\\
  &=&\frac{1}{2^N} \sum^{2^N-1}_{k_A=0}\sum^{2^N-1}_{k_B=0}\vert \varphi_{k_A}\rangle_{A_1A_2...A_N}\vert \varphi_{k_B}\rangle_{B_1B_2...B_N}\otimes\nonumber\\&&
  [\frac{1}{(\sqrt{2})^N}\sum^1_{l_1,...,l_N=0}\exp(\frac{-2\pi ij(k_A+k_B)}{4})\exp(i\delta_j))\vert l_N,l_{N-1},...,l_1\rangle_{C_1C_2...C_N}].\nonumber\\
\end{eqnarray}
We find that no matter what measurement results Alice and Bob obtain, the state of Charlie's particles can always be transformed to the desired form via unitary operations:
\begin{eqnarray}
U_{k_Ak_B}&=&\sum^1_{l_1,...,l_N=0}\exp(\frac{2\pi ij(k_A+k_B)}{4})\vert l_N,l_{N-1},...,l_1\rangle\langle l_N,l_{N-1},...,l_1\vert\nonumber\\
&=&\prod^\otimes \sum^1_{l_n}\exp(\frac{2\pi i\times2^{n-1}l_n}{4})\vert l_n\rangle\langle l_n\vert.
\end{eqnarray}
Charlie only has to perform $N$ single-qubit unitary operations based on Alice's and Bob's measurement results $k_A$ and $k_B$ in order to obtain the state.

\section{Discussion}
We have revisited the scheme described in Ref.\cite{JRSP2E} for JRSP of two-qubit equatorial states. We have shown that  the success probability of the scheme can be improved from 25\% to 100\% by choosing optimal measurement bases for the two senders . We have also successfully generalized our scheme to remotely prepare $N$-qubit equatorial states in a deterministic manner.

In our scheme, there are only two senders. Actually, the scheme can be simply changed to a JRSP protocol with $M$ senders. Suppose each sender holds partial information where $a_j+b_j+...+m_j=\delta_j$ $(j=0,1,...2^N-1)$. In this case the quantum channel should be $N$ $(M+1)$-qubit GHZ states. Each of the senders measures her/his own $N$ particles in an appropriate basis $\vert \varphi\rangle$ based on her/his information. Then the receiver can obtain the desired state via unitary operations conditioned on the $M$ senders' measurement results.
Moreover, the scheme can also be generalized to a controlled RSP (CRSP) scheme \cite{MCRSP_I_NE_EPR,CRSP_II_III_B,CJRSP_I,wc,ANCRSP} by increasing the number of qubits in the GHZ states. Generally, the controller has no information about the quantum state to be prepared. Therefore, he/she only needs to perform $N$ single-particle measurements in the diagonal basis $\vert \pm\rangle=\frac{1}{\sqrt{2}}(\vert 0\rangle\pm\vert 1\rangle)$.

The equatorial states have some special properties that make them interesting for quantum information processing \cite{eq1,eq2,eq3}. Since they contain less information compared to arbitrary quantum states, it should be easier to prepare equatorial states than arbitrary states. It was previously shown that a single-qubit equatorial state can be remotely prepared with one classical bit via the maximally entangled channel \cite{RSP_pati}. From our results, we can conjecture that the equatorial state can always be deterministically prepared via a proper maximally entangled channel.


\begin{thebibliography}{99}
\bibitem{RSP_lo} Lo, H.K.: Classical-communication cost in distributed quantum-information processing: a generalization
of quantum-communication complexity. Phys. Rev. A 62, 012313 (2000)

\bibitem{tele} Bennett, C.H., Brassard, G., Cr¨¦peau, C., Jozsa, R., Peres, A.,Wootters, W.K.:Teleporting an unknown quantum state via dual classical and Einstein-Podolsky-Rosen channels. Phys. Rev. Lett. 70, 1895-1899 (1993)

\bibitem{RSP_Bennet} Bennett, C.H., DiVincenzo, D.P., Shor, P.W., Smolin, J.A., Terhal, B.M., Wootters,W.K.: Remote state
preparation. Phys. Rev. Lett. 87(7), 077902 (2001)

\bibitem{RSP_pati} Pati, A.K.: Minimum classical bit for remote preparation and measurement of a qubit. Phys. Rev. A
63, 014302 (2000)

\bibitem{low_entanglement_RSP} Devetak, I., Berger, T.: Low-Entanglement Remote State Preparation. Phys. Rev. Lett. 87, 197901 (2001)

\bibitem{optimal_RSP} Berry, D.W., Sanders, B.C.: Optimal remote state preparation. Phys. Rev. Lett. 90(5), 057901 (2003)

\bibitem{generalized_RSP} Abeyesinghe, A., Hayden, P.: Generalized remote state preparation: Trading cbits, qubits, and ebits in quantum communication. Phys. Rev. A 68, 062319 (2003)

\bibitem{oblivious_RSP} Leung, D.W., Shor, P.W.: Oblivious remote state preparation. Phys. Rev. Lett. 90(12), 127905 (2003)

\bibitem{faithful_RSP} Ye, M.Y., Zhang, Y.S., Guo, G.C.: Faithful remote state preparation using finite classical bits and a nonmaximally entangled state. Phys. Rev. A 69, 022310 (2004)

\bibitem{cv_RSP} Paris, M.G.A., Cola, M.M., Bonifacio, R.:  Remote state preparation and teleportation in phase space. J. Opt. B 5, S360 (2003)

\bibitem{RSPE1} Peng, X.H., Zhu, X.W., Fang, X.M., Feng, M., Liu,M.L., Gao, K.L.: Experimental implementation of
remote state preparation by nuclear magnetic resonance. Phys. Lett. A 306, 271-276 (2003)

\bibitem{RSPE2} Xiang, G.Y., Li, J., Bo, Y., Guo, G.C.: Remote preparation of mixed states via noisy entanglement.
Phys. Rev. A 72(1), 012315 (2005)

\bibitem{RSPE3} Peters, N.A., Barreiro, J.T., Goggin, M.E., Wei, T.C., Kwiat, P.G.: Remote State Preparation: arbitrary remote control of photon polarization. Phys.
Rev. Lett. 94(14), 150502 (2005)

\bibitem{RSPE4} Liu, W.T., Wu, W., Ou, B.Q., Chen, P.X., Li, C.Z., Yuan, J.M.: Experimental remote preparation of
arbitrary photon polarization states. Phys. Rev. A 76(2), 022308 (2007)

\bibitem{RSPE5} Rosenfeld, W., Berner, S., Volz, J., Weber, M., Weinfurter, H.: Remote Preparation of an Atomic Quantum Memory. Phys. Rev. Lett. 98, 050504 (2007)

\bibitem{RSPE6} Barreiro, J.T., Wei, T.C., Kwiat. P.G.: Remote Preparation of Single-Photon ¡°Hybrid¡± Entangled and Vector-Polarization States. Phys. Rev. Lett. 105, 030407 (2010)



\bibitem{JRSP_I_GC1} Xia, Y., Song, J., Song, H.S.: Multiparty remote state preparation. J. Phys. B: At. Mol. Opt. Phys.
40(18), 3719-3724 (2007)

\bibitem{JRSP_I_GC2} An, N.B., Kim, J.: Joint remote state preparation. J. Phys. B At. Mol. Opt. Phys. 41(9), 095501 (2008)

\bibitem{AN0} An, N.B.: Joint remote preparation of a general two-qubit state. J. Phys. B 42, 125501 (2009)

\bibitem{JRSP1} Zhan, Y.B., Zhang, Q.Y., Shi, J.: Probabilistic joint remote preparation of a high-dimensional equatorial quantum state. Chin. Phys. B 19, 080310 (2010)

\bibitem{JRSP2} Luo, M.X., Chen, X.B., Ma, S.Y., Niu, X.X., Yang, Y.X.: Joint remote preparation of an arbitrary
three-qubit state. Opt. Commun. 283(23), 4796-4801 (2010)

\bibitem{JRSP3} Chen, Q.Q., Xia, Y., An, N.B.: Joint remote preparation of an arbitrary three-qubit state via EPR-type
pairs. Opt. Commun. 284, 2617-2621 (2011)

\bibitem{AN1} An, N.B., Kim, J.: Collective remote state preparation. Int. J. Quantum Inform. 6, 1051-1066 (2008)

\bibitem{AN2} An, N.B.: Joint remote state preparation via W and W-type states. Opt. Commun. 283, 4113-4117 (2010)

\bibitem{AN3} Chen, Q.Q., Xia, Y., Song, J., An, N.B.: Joint remote state preparation of a W-type state via W-type states. Physics Letters A 374, 4483-4487(2010)

\bibitem{Xiao} Xiao, X.Q., Liu, J.M., Zeng, G.: Joint remote state preparation of arbitrary two- and three-qubit states.
J. Phys. B 44, 075501 (2011)

\bibitem{Peng} Peng, J.Y., Luo, M.X., Mo, Z.W.: Joint remote state preparation of arbitrary two-particle states via
GHZ-type states. Quantum Inf. Process. 12, 2325¨C2342 (2013)

\bibitem{JRSP2E} Choudhury, B.S., Dhara, A.:¡¡Joint remote state preparation for two-qubit¡¡equatorial states£®Quantum Inf. Process. 14, 373¨C379 (2015)¡¡



\bibitem{MCRSP_I_NE_EPR} Wang, Z.Y., Liu, Y.M., Zuo, X.Q., Zhang, Z.J.: Controlled remote state preparation. Commun. Theor.
Phys. (Beijing, China) 52(2), 235-240 (2009)





\bibitem{CRSP_II_III_B} Chen, X.B., Ma, S.Y., Yuan, S.Y., Zhang, R., Yang, Y.X.: Controlled remote state preparation of arbitrary two and three qubit states via the Brown state. Quantum Inf. process. 11, 1653-1667 (2012)






\bibitem{CJRSP_I} Wang, D., Ye, L.: Multi party-controlled joint remote preparation. Quantum Inf. process. 12, 3223-3237
(2013)

\bibitem{wc} Wang, C., Zeng, Z., Li, X.H.: Controlled remote state preparation via partially
entangled quantum channel. Quantum Inf. process. 14, 1077-1089
(2015)

\bibitem{ANCRSP} An, N.B., Bich, C. T.: Perfect controlled joint remote state preparation independent ofentanglement degree ofthequantum channel. Phys. Lett. A 378, 3582-3585 (2014)

\bibitem{eq1} Buscemi, F.,D¡¯Ariano, G.M., Macchiavello, C.: economical phase-covariant cloning of qudits. Phys.
Rev. A 71, 042327 (2005)
\bibitem{eq2} Rezakhani, A.T., Siadatnejad, S., Ghaderi, A.H.: separability in asymmetric phase-covariant cloning.
Phys. Lett. A 336, 278 (2005)
\bibitem{eq3} Li, X.H., Ghose, S.: Control power in perfect controlled teleportation via partially entangled channels. Phys. Rev. A 90, 052305 (2014)

\end{thebibliography}
\end{document}